\documentclass[aps,prl,twocolumn,amsmath,amssymb]{revtex4}

\usepackage{graphicx}
\usepackage{dcolumn}
\usepackage{bm}
\usepackage[caption=false]{subfig}
\usepackage{amssymb}
\usepackage{amsmath}
\usepackage{graphicx}
\usepackage{verbatim}
\usepackage[usenames, dvipsnames]{color}
\usepackage{hyperref}
\usepackage{color}

\begin{document}

\title{Fermi-liquid behavior and thermal conductivity of $\epsilon$-iron at Earth's core conditions}

\author{L. V. Pourovskii$^{1,3}$, J. Mravlje$^2$, A. Georges$^{3,1,4}$, S.I. Simak$^5$, I. A. Abrikosov$^{5,6}$}

\affiliation{
$^1$Centre de Physique Th{\'e}orique, Ecole Polytechnique, CNRS, Universit{\'e} Paris-Saclay, F-91128 Palaiseau, France \\
$^2$Jozef Stefan Institute, SI-1000, Ljubljana, Slovenija \\
$^3$ Coll{\`e}ge de France, 11 place Marcelin Berthelot, 75005 Paris, France \\
$^4$Department of Quantum Matter Physics, University of Geneva, 24 Quai Ernest-Ansermet, 1211 Geneva 4, Switzerland \\
$^5$Department of Physics, Chemistry and Biology (IFM), Link\"oping University, SE-58183 Link\"oping, Sweden \\
$^6$Materials Modeling and Development Laboratory, National University of Science and Technology "MISIS", Moscow, Russia}

\date{\today}

\begin{abstract} 

The electronic state and transport properties of hot dense iron are of the utmost importance to geophysics.  Combining the density functional and
dynamical mean field theories we study the impact of electron
correlations on electrical and thermal resistivity of hexagonal
close-packed $\epsilon$-Fe at Earth's core conditions.  $\epsilon$-Fe
is found to behave as a nearly perfect Fermi liquid. The quadratic
dependence of the scattering rate in Fermi liquids leads to a
modification of the Wiedemann-Franz law with suppression of the
thermal conductivity as compared to the electrical one. This
significantly increases the electron-electron thermal resistivity
which is found to be of comparable magnitude to the electron-phonon
one. The implications of this effect on the dynamics of Earth's core
is discussed.

\end{abstract}
 
\maketitle

Earth's magnetic field plays a crucial role in the survival of the
human race. It keeps the ozone layer intact despite the solar wind and
therefore protects the Earth from destructive ultraviolet
radiation\cite{Tarduno2015}. The magnetic field is generated by
self-sustained dynamo action in its iron-rich
core\cite{Olson2013}. This geodynamo runs on heat from the growing
solid inner core and on chemical convection provided by light elements
issued from the liquid outer core on
solidification\cite{Pozzo2012}. The power supplied to drive the
geodynamo is proportional to the rate of inner core growth, which in
turn is controlled by heat flow at the core-mantle
boundary\cite{Lay2008}.  This heat flow critically depends on the
thermal and electrical conductivities of liquid iron under the extreme
pressure and temperature conditions in the Earth's core. For a long
time there has been agreement that convection in the liquid outer core
provides most of the energy for the geodynamo and does so for at least
3.4 billion years\cite{Olson2013,Stacey2007}.

Recently, such a view has been challenged by first-principles
calculations\cite{deKoker2012,Pozzo2012}, suggesting a much higher
capacity for the liquid core to transport heat by conduction and
therefore less ability to transport heat by
convection\cite{Olson2013}.  The calculated conductivities have been
found to be two to three times higher than the generally accepted
estimates, urging for reassessment of the core thermal history and
power requirements\cite{Pozzo2012}.

Convection also plays a crucial role in the current theory of the
solid core dynamics, as a radial motion of the inner core matter is
invoked to explain the observed seismic anisotropies of the inner
core\cite{Romanowicz1996,Buffett2009,Monnereau2010}. However, {\it ab initio}
calculations\cite{Pozzo2014} similarly predict a too high thermal
conductivity for hexagonal close-packed (hcp) $\epsilon$-iron
generally assumed to form the inner core, thus impeding a significant
convection of its solid matter.  The first-principles calculations for
liquid and solid iron of Refs.\cite{deKoker2012,Pozzo2012,Pozzo2014},
unlike previous results, have not relied on any extrapolations,
however, they employed the standard density-functional-theory (DFT)
framework in which dynamical many-body effects are neglected.

Many-body effects in crystalline iron at the conditions of Earth's
inner core have been previously studied in
Refs.~\cite{pourovskii2013,Vekilova2015} using the density functional
theory plus dynamical mean-field theory (DFT+DMFT)
method\cite{georges_dmft_1996,kotliar_elec_struc_dmft_2006}.  The hcp
$\epsilon$-phase was predicted to exhibit a typical Fermi liquid
behavior with a quadratic temperature dependence of the
electron-electron scattering rate, $\Gamma$. In contrast, the
body-centered cubic (bcc) $\alpha$-phase at the same conditions was
shown to feature a strongly non-Fermi-liquid electron-electron
scattering.

    \begin{figure*}
    	\begin{center}
    		\includegraphics[width=1.9\columnwidth]{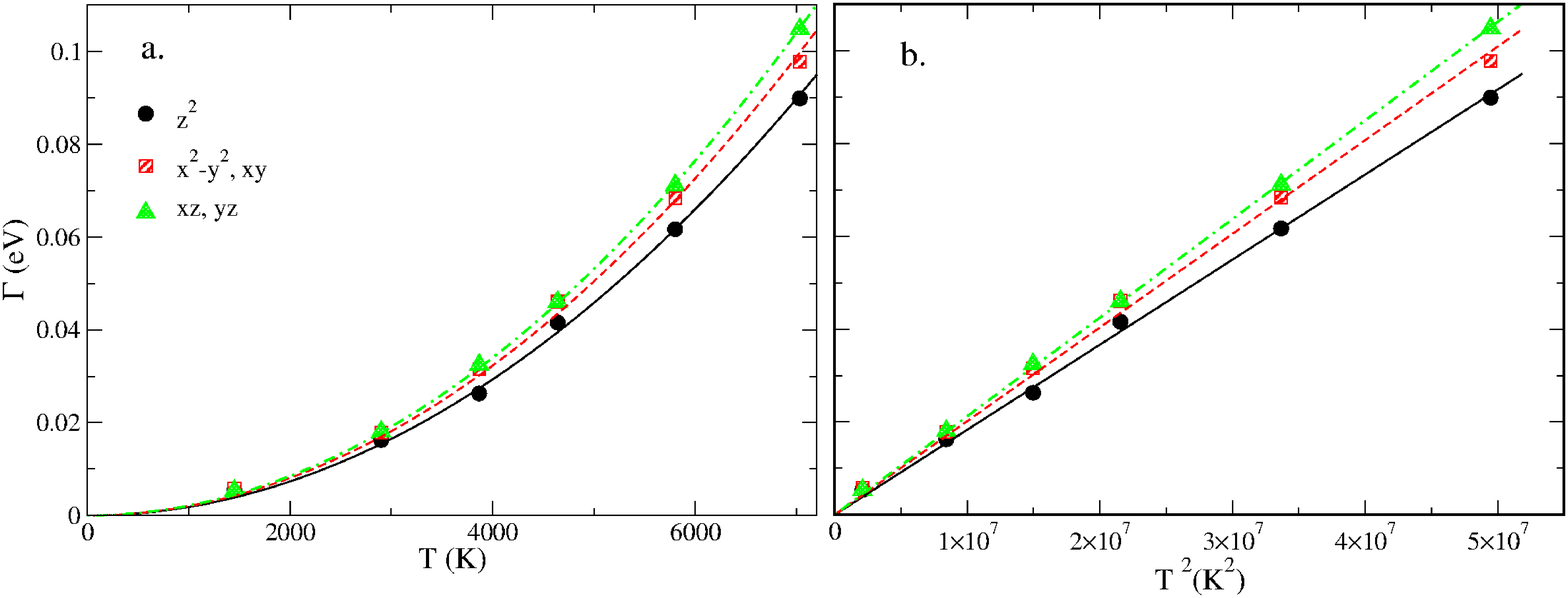}
    	\end{center}
    	\caption{\label{fig:Gamma} (Color online)  Orbitally-resolved values of the inverse quasi-particle lifetime $\Gamma$ in $\epsilon$-Fe as a
    		function of temperature T. a. $\Gamma$ vs. $T$. The curves are fits $AT^2$ to the calculated data. b. $\Gamma$ vs. $T^2$. The
    		curves are linear-regression fits to the data. The values of $\Gamma$ were extracted from real-axis self-energies
    		obtained by the Maximum-entropy method \cite{Beach2004}.
    	}
    \end{figure*}

Zhang et al.\cite{Zhang2015} have later pointed out that the effect of electron-electron scattering (EES) of  $d$-electrons due to correlations  is missing in previous transport calculations within DFT~\cite{Pozzo2012,Pozzo2014} and treatment of electron-phonon scattering (EPS) only is not sufficient.   Using the same DFT+DMFT method they predicted a non-Fermi-liquid linear temperature dependence of the ESS in compressed $\epsilon$-iron, in disagreement with the conclusions of Ref.~\cite{pourovskii2013}. The EES contribution to the electrical resistivity at core temperatures was predicted to be as large as the electron-phonon one \cite{Zhang2015}.   This principal result was later retracted \cite{Zhang2016} because of a numerical mistake in their transport calculations leading to a significant  overestimation of the EES electrical resistivity.

Obviously, it is of the utmost importance to Earth's physics to clearly
elucidate how large the EES contribution to the electrical and thermal
resistivity at Earth's core conditions is. This is the motivation and
the main subject of the present letter. We perform a detailed and precise
calculation of the quasiparticle (QP) properties, especially of the QP
scattering rate, and establish  the FL nature of $\epsilon$-iron at the
inner core conditions. Most importantly, the quadratic frequency dependence of the scattering rate characteristic of Fermi liquids 
has a direct bearing on the transport properties of $\epsilon$-Fe, as
demonstrated here by an explicit calculation of the electrical and
thermal conductivity. In Fermi liquids, the  Lorenz number in
the Wiedemann-Franz law is suppressed, thus the EES contribution to
the thermal resistivity is enhanced. The EES contribution to the thermal
resistivity is of comparable magnitude to the EPS one and should not
be neglected. By including both contributions we obtain a
substantially reduced value for the total thermal conductivity of pure
$\epsilon$-Fe at the inner core conditions as compared to previous DFT
calculations\cite{Pozzo2014}. Hence, the Fermi-liquid nature of
$\epsilon$-Fe suppresses its thermal conductivity and may play an
important role in stabilizing the convection in the Earth core.

    We employed the self-consistent DFT+DMFT implementation\cite{Aichhorn2009,Aichhorn2011,triqs_dft_tools} in a full-potential framework\cite{blaha2001wien2k}. We used the same parameters as in Ref.~\cite{pourovskii2013} for the lattice (volume 7.05 \AA$^3$/atom, the hcp c/a ratio 1.6) and construction of the Wannier orbitals (energy window [10.8 eV, 4.0 eV] around the Fermi level), as well as around-mean-field double counting. The rotationally-invariant Coulomb interaction was defined by the parameters F0=U=5.0 eV and J=0.93 eV. The DMFT quantum impurity problem was solved using the hybridization-expansion continuous-time quantum Monte-Carlo method\cite{Gull2011} as implemented in Ref.~\cite{Seth2016}.  The same parameters were used for both hcp and bcc Fe. 
    For the analytical continuation we employed the Maximum-entropy (MaxEnt) method in the implementation of Ref.~\cite{Beach2004}. 
    The conductivity was calculated as described in Refs.~\cite{triqs_dft_tools} and \cite{pourovskii2014}. 

    \begin{figure}
    	\begin{center}
    		\includegraphics[width=0.9\columnwidth]{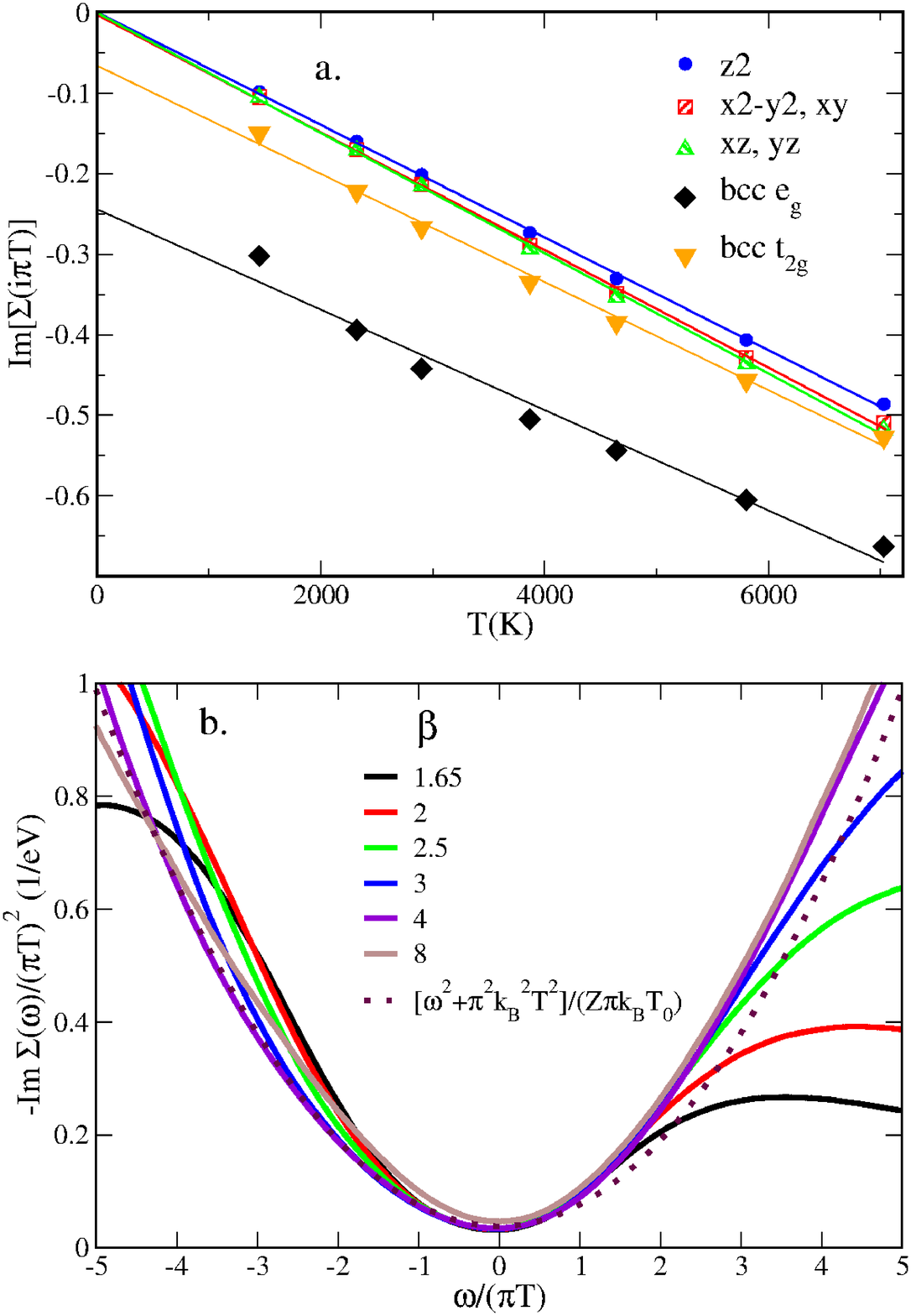}
    	\end{center}
    	\caption{\label{fig:FL} (Color online) Fermi-liquid scaling of the DMFT self-energy in $\epsilon$-Fe. a. The imaginary part of the DMFT self-energy at the first Matsubara point $\omega_1=i\pi k_B T$ vs. temperature for hcp and bcc
Fe. Note that $Im[\Sigma(i\pi k_B T)]$ being proportional to T is a signature of a Fermi-liquid\cite{Chubukov2012} . The
lines are the linear regression fits to the calculated points for corresponding 3$d$ orbitals of Fe.
b. The rescaled imaginary part of the DMFT self-energy at the real axis, $Im[\Sigma(\omega)]/(\pi k_B T)^2$ ,
vs. $\omega/(\pi k_B T)$. The real-frequency self-energies are obtained by the MaxEnt analytic continuation
method\cite{Beach2004}. One sees that all self-energies collapse into a single curve described by a parabolic
fit (the dotted line) defined by the quasiparticle weight $Z=$0.7 and the characteristic Fermi-liquid
temperature scale $T_0=$12~eV. Deviations from the Fermi-liquid behavior of the
resistivity are expected for temperatures above $T_{FL}\sim$0.1~$T_0$ , see Ref.~\cite{Berthod2013}, i.e. $T_{FL}$ corresponds to about 14~000~K in the present case.
    	}
    \end{figure} 

First we analyze the temperature dependence of the inverse quasiparticle life-time $\Gamma_m=-Z_m Im [\Sigma_m(\omega=0)]$, where $\Sigma_m(\omega=0)$ is the value of DMFT self-energy for the orbital $m$ at zero frequency, $Z_m$ is the corresponding quasi-particle residue, $Z_m=\left(1-\left.\frac{d Re [\Sigma_m(\omega)]}{d\omega}\right|_{\omega \to 0}\right)^{-1}$.
Our resulting dependence of $\Gamma$ vs. $T$ is plotted in Fig.~\ref{fig:Gamma}a One may notice a clearly parabolic Fermi-liquid shape of $\Gamma(T)$ for all three inequivalent orbitals of the 3$d$ shell of Fe in the hcp lattice. Correspondingly, $\Gamma$  scales linearly as a function of $T^2$, see  Fig.~\ref{fig:Gamma}b. In contrast, the values of $\Gamma$ obtained by Zhang et al.~\cite{Zhang2015} exhibit a non-Fermi-liquid linear
dependence on $T$. While our values agree with theirs at T=6000 K, for lower T the difference
is significant. To obtain $\Gamma$ plotted in Fig. 1 we 
have analytically continued the imaginary-frequency  DMFT  self-energy $\Sigma(i\omega_n)$, where $\omega_n$ is the fermionic Matsubara frequency $\omega_n=\pi(2n-1)k_BT$ and $k_B$ is the Boltzmann constant,  to the real-frequency axis using the MaxEnt. 

Our results for  the scattering rate $\Gamma$ shown  in Fig.~\ref{fig:Gamma} are obtained from analytically-continued DMFT self-energy. It is well known that the analytical continuation methods needed to obtain the real-frequency data from the imaginary-frequency self-energy are quite sensitive to the details of the procedure (e.g. the number of Matsubara
frequencies included into the Pade approximant\cite{Beach2000}, the way high frequency noisy tails are treated and the way the stochastic error is estimated in the initial imaginary-time data in the case of the MaxEnt etc.). However, a qualitative but definite conclusion about the Fermi or non-Fermi-liquid nature of a system can be  inferred directly from the imaginary-frequency self-energy without resorting to any analytical continuation. This is done by employing the so-called "first-Matsubara-frequency" rule. As demonstrated, e. g., in Ref.~\cite{Chubukov2012}, in a Fermi liquid the imaginary
part of electronic self-energy, $\Sigma$, at the first Matsubara point within a local approximation like
DMFT must be proportional to the temperature, T, i.e.  $Im[\Sigma(i\pi k_B T)]=\lambda T$, where $\lambda$ is a real
constant. In Fig.~\ref{fig:FL}a we plot $Im[\Sigma(i\pi k_B T)]$ as a function of temperature for all
inequivalent orbitals in hcp and bcc Fe. One may clearly see that in the $\epsilon$ phase $Im[\Sigma(i\pi k_B T)]$ is almost perfectly proportional to T, in contrast to bcc Fe, where it exhibits significant
deviations from the "first-Matsubara-frequency" rule\cite{pourovskii2013}. This result confirms the Fermi-liquid state of $\epsilon$-Fe at Earth's core conditions. We note that this conclusion is further corroborated by a weak temperature dependence of the
of our calculated quasiparticle weight $Z$, as well as by the ratio $\frac{\Gamma}{T} \ll 1$.

Moreover, our real-frequency self-energies for different temperatures collapse into a curve consistent with the dependence $const \cdot (\omega^2+(\pi k_B T)^2)$ expected for a Fermi liquid, see Fig.~\ref{fig:FL}b . From this plot we extracted the upper bound $T_{FL}$ for the Fermi-liquid regime of the transport following Ref.~\cite{Berthod2013}. The obtained temperature $T_{FL}\approx$ 14000~K is
much higher than temperatures expected for Earth's core.

\begin{figure}
    	\begin{center}
    		\includegraphics[width=0.9\columnwidth]{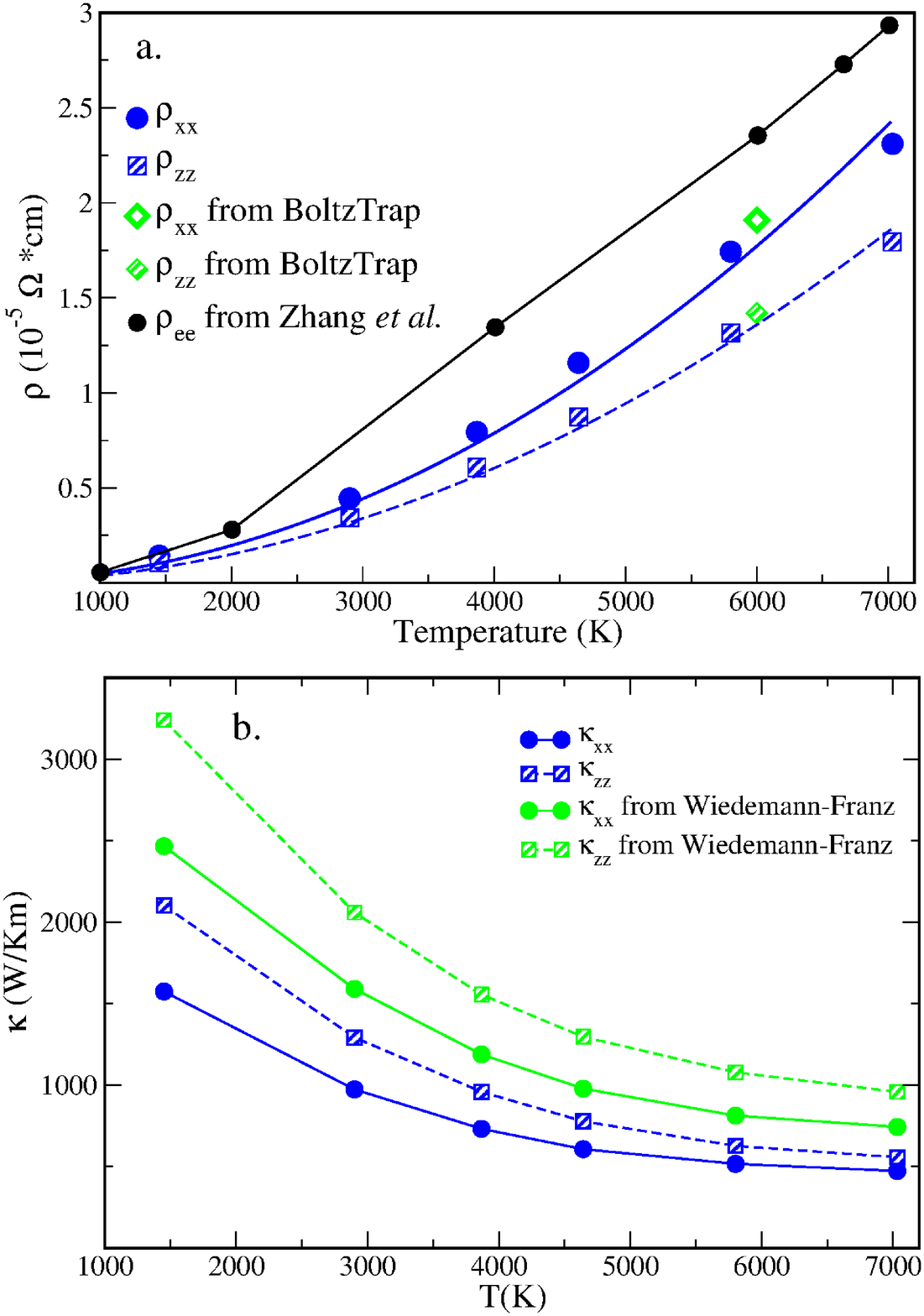}
    	\end{center}
    	\caption{\label{fig:conduct} (Color online)Calculated electron-electron-scattering contribution to the electrical and thermal
    		resistivity of hcp iron at Earth's core density. a. Electrical resistivity. Blue filled circles and hashed squares are our DFT+DMFT results for $\rho_{xx}$ and $\rho_{zz}$ , respectively. Red filled and hashed triangles are the corresponding resistivities calculated by us with the imaginary part of
    		the self-energy from Extended Data Fig. 3 of Zhang et al.~\cite{Zhang2015}. Green empty and hashed diamonds are the corresponding resistivities calculated by the Boltzmann-transport code BoltzTrap \cite{Madsen2006} assuming a
    		Fermi-liquid with $\Gamma/Z=$0.09~eV. Black circles are the electron-electron-scattering contribution
    		to the electrical resistivity reported obtained Zhang et al.~\cite{Zhang2015} including the correction for the omitted spin factor \cite{Zhang2016}. b. Thermal conductivity. Blue
    		filled circles and hashed squares are DFT+DMFT results for $\kappa_{xx}$ and $\kappa_{zz}$, respectively. The
    		green lines/symbols are the corresponding conductivities obtained from our calculated
    		electrical conductivity using the Wiedemann-Franz law with the standard Lorenz number of 2.44$\cdot$10$^{-8}$~W$\Omega$K$^{-2}$.
    	}
    \end{figure}

Fig.~\ref{fig:conduct}a shows our calculated contribution of the electron-electron scattering to the electrical resistivity. One clearly observes that it increases quadratically with increasing temperature up to at least 6000 K, corresponding to Earths core conditions, in contrast to the results of Ref.~\cite{Zhang2015}. The obtained value of about 1.6$\cdot$ 10$^{-5}$~$\Omega \cdot$cm at T=6000~K is rather insignificant compared to the electron-phonon-scattering contribution of about 5.3$\cdot$10$^{-5}$~$\Omega \cdot$cm predicted by DFT calculations \cite{Pozzo2014}   indicating that the electron-electron
scattering cannot strongly influence the electrical resistivity in hcp-Fe at Earth's core conditions. 

In Fig.~\ref{fig:conduct}b we display the corresponding thermal conductivity due to electron-electron scattering. 
One may notice that this conductivity is not very high: its average magnitude of 540~Wm$^{-1}$K$^{-1}$ at 6000~K is comparable to the figure  
$\sim 300$~Wm$^{-1}$K$^{-1}$ obtained in Ref.~\cite{Pozzo2014} for the electron-phonon thermal conductivity. 
By including both scattering effects the total conductivity is reduced to about 190~Wm$^{-1}$K$^{-1}$, hence, the corresponding resistivity is
enhanced by about 60\%. 

In fact, this large electron-electron-scattering contribution is directly related to the Fermi-liquid behavior of $\epsilon$-Fe.
One may demonstrate this by simple analytical calculations~\cite{herring1967,herring_erratum,supp_kappaFL}.  
Using a Fermi-liquid scattering rate $1/\tau(\omega)=1/\tau_0\cdot\left[1+\omega^2/(\pi k_B T)^2\right]$ 
(with $1/\tau_0\propto T^2$), the electrical conductivity ($\sigma$) and thermal conductivity ($\kappa$) 
are found to be, in the low-temperature limit $T\lesssim T_{FL}$~\cite{supp_kappaFL}:
\begin{equation}
\frac{\sigma}{\sigma_0}\,=\,\frac{I_{01}}{I_{00}}\simeq 0.82
\,\,\,,\,\,\,
\frac{\kappa}{\kappa_0}\,=\,\frac{I_{21}}{I_{20}}\simeq 0.53
\end{equation}
in which $\sigma_0$ and $\kappa_0$ denote the conductivities obtained with the frequency-independent scattering time $\tau_0$. 
In these expressions,  
$I_{nk}\equiv \int_{-\infty}^{+\infty}dx\, x^n (1+x^2/\pi^2)^{-k}\cosh^{-2}(x/2)$ are transport integrals and $x=\omega/T$. 
Hence, the Lorenz number for such a Fermi-liquid with inelastic scattering only is equal to \cite{herring1967,herring_erratum}:
$$
\frac{\kappa/T}{\sigma}=L_{FL}=L_0/1.54
$$
where $L_0=\pi^2/3 (k_B/e)^2$ is the conventional Lorenz number for a frequency-independent scattering rate. 
The stronger effect of the frequency-dependence of $\tau(\omega)$ on the thermal conductivity as compared to $\sigma$ 
is due to the additional power $x^2$ in the numerator of the transport integrals for $\kappa$. 
Using the conventional value $L_0$ of the Lorenz number together with our calculated $\sigma$
would lead one to a substantially larger thermal conductivity (see Fig.~\ref{fig:conduct}b), and hence an incorrect conclusion 
that the electron-electron contribution to the thermal scattering is insignificant, too. 
This calculation can be generalized to take into account other sources of scattering on top of purely inelastic EES, 
such as impurity or electron-phonon scattering, leading to a $T$-dependent Lorenz number, as detailed in the supplemental material. 

In conclusion, we have established $\epsilon$- iron is a Fermi liquid
at Earth's core conditions. We have shown that implications of this
finding are far reaching as the electron-electron inelastic scattering
characteristic of Fermi-liquids significantly suppresses the thermal
conductivity of Earth's inner core. The quadratic frequency dependence
of this scattering leads to a reduction of the Lorenz number, hence,
the thermal conductivity is suppressed with respect to predictions of
the conventional Wiedemann-Franz law. As a result, the
electron-electron-scattering contribution is comparable to the
electron-phonon one. By taking them both into account, we obtained a
significant reduction of the thermal conductivity of the $\epsilon$
phase at the inner core's condition, which supports explanations of
the inner core anisotropy in terms of convection processes.

 The same effects may be important for liquid iron, too.  The obtained
 reduction is insufficient to explain the stability of convection by
 itself. But it is likely that the thermal disorder and the
 admixture of significant quantities of light elements
 \cite{ORourke2016}, that we did not take into account may further decrease the thermal conductivity. The impact of alloying, crystalline order and thermal
 vibrations on  electronic correlations should be investigated in future work. Finally, we note
 that the long-wave length spin-fluctuations that are disregarded in
 our approach may lead to additional suppression of the Lorenz number \cite{rice}.

{\it Acknowledgments} L.V. P. acknowledges the financial support of
the Ministry of Education and Science of the Russian Federation in the framework of
Increase Competitiveness Program of NUST MISiS (No. K3-2015-038). J. M. is supported by the Slovenian Research Agency (ARRS)  under Program P1-0044. A.G., J.M. and L.P. acknowledge the support of the European Research Council grant ERC-319286 QMAC.
 S.I.S. and I.A.A. acknowledge the Swedish Research Council (VR) Projects No. 2014-4750 and 2015-04391, LiLi-NFM, and the Swedish Government Strategic Research Area in Materials Science on Functional Materials at Link\"oping University (Faculty Grant SFO-Mat-LiU No. 2009 00971). I.A.A. is grateful for the support provided by the Swedish Foundation for Strategic Research (SSF) program SRL Grant No. 10-0026, as well as by the Ministry of Education
and Science of the Russian Federation (Grant No. 14.Y26.31.0005). 
The computations were performed on resources provided by CPHT-Ecole Polytechnique as well as by the Swedish National Infrastructure for Computing (SNIC)  at National Supercomputer Centre (NSC) and Center for High Performance Computing (PDC).


\newpage
\onecolumngrid
\setcounter{equation}{0}
\renewcommand\theequation{S\arabic{equation}}
\setcounter{figure}{0}
\renewcommand\thefigure{S\arabic{figure}}

\centerline{\LARGE{Supplemental material}}
\vspace{5mm}

In a standard Boltzmann formalism within the relaxation time approximation the conductivity
($\sigma$) and the thermal conductivity ($\kappa$) are given as specified, e.g., in Ref.~\cite{AshcroftMermin}:

\begin{equation}\label{eq:sig}
	\sigma=e^2\int d\epsilon\Phi(\epsilon)(-f'(\epsilon))\tau(\epsilon),
\end{equation}
\begin{equation}\label{eq:kap}
\kappa=\frac{1}{T}\int d\epsilon^2\Phi(\epsilon)(-f'(\epsilon))\tau(\epsilon)-\frac{\left[\int d\epsilon\epsilon\Phi(\epsilon)(-f'(\epsilon))\tau(\epsilon)\right]^2}{\int d\epsilon\Phi(\epsilon)(-f'(\epsilon))\tau(\epsilon)}
\end{equation}
where $\epsilon$ is the energy measured with respect to the chemical potential, $\Phi(\epsilon)$ is the transport
function, $f$ is the Fermi function and $\tau$ is the relaxation time. Often the energy dependence of
$\tau$ is neglected. If one additionally neglects the energy dependence of transport function and
evaluates the elementary integrals, one gets the Wiedemann-Franz law
$$
\kappa/(\sigma T)=\frac{\pi^2}{3}\left(\frac{k_B}{e}\right)^2=L,
$$
where the Lorenz number $L$ is 2.44$\cdot$10$^{-8}$W$\Omega$K$^{-2}$.

In the case of a Fermi liquid, however, the energy dependence of scattering rate is very strong
$$
1/\tau(\epsilon)=1/\tau(\epsilon=0)\cdot(1+\epsilon^2/(\pi T k_B)^2)
$$

This leads to a modification of the Wiedemann-Franz law
$$
\kappa/(\sigma T)=L/1.54=L_{FL}
$$

Accordingly, the conductivity and thermal conductivity from Eqs.~\ref{eq:sig} and \ref{eq:kap} will be smaller
by 0.82 and 0.53, respectively, if compared with that obtained by neglecting the energy
dependence of the scattering rate, i.e. by putting
$$
1/\tau(\epsilon)=1/\tau(\epsilon=0)
$$
into Eqs.~\ref{eq:sig} and \ref{eq:kap} .

Remarkable suppression of the thermal conductivity is especially important for the discussion
in the main text.

One may estimate the impact of this effect on the overall thermal conductivity of $\epsilon$-Fe by
summing up the contributions from electron-electron and electron-phonon scattering. To
obtain the later we evaluated the ratio $\kappa/\tau$ using the BoltzTraP\cite{Madsen2006} code. By adopting for the
conductivity with electron-phonon scattering, $\kappa_{e-ph}$ , the value of 300 Wm$^{-1}$K$^{-1}$ obtained by DFT calculations of Ref.~\cite{Pozzo2014} , we estimated the electron-phonon quasiparticle lifetime $\tau_{e-ph}=$1.11$\cdot$10$^{-15}$ s. Assuming a frequency-independent electron-phonon scattering one obtains for the total lifetime :
$$
\frac{1}{\tau_{tot}}=\frac{1}{\tau_{e-ph}}\left[1+\frac{\tau_{e-ph}}{\tau(\epsilon=0)}(1+\epsilon^2/(\pi T k_B)^2)\right].
$$

The electron-electron scattering contribution to the thermal conductivity evaluated with full
DMFT transport calculations is shown in Fig.~\ref{fig:conduct}b of the main text. Here we present a simple
semi-classical calculations, where electron-electron-scattering lifetime is obtained from the
average value of self-energy at zero frequency, 0.09 eV. Hence, $\tau(\epsilon=0)=\hbar/(2\Sigma(0))$ is
3.66$\cdot$10$^{-15}$ s and the ratio $\tau_{e-ph}/\tau(\epsilon=0)$=0.303. Inserting $\tau_{tot}$ into (\ref{eq:kap}) and carrying out the integration assuming a constant value for the transport function one obtains the reduction of $\kappa$ by a factor of 0.61 as compared to pure electron-phonon scattering. Hence, the thermal conductivity is reduced from 300 to 183~Wm$^{-1}$K$^{-1}$. This value is very
close to the one obtained by adding the electron-electron thermal scattering calculated directly
within DMFT (Fig~\ref{fig:conduct}.b of the main text) to the electron-phonon contribution, $\kappa_{tot}=1/(\kappa_{e-e}^{-1}+\kappa_{e-ph}^{-1})=$190~Wm$^{-1}$K$^{-1}$.

\begin{figure}
 	\begin{center}
 		\includegraphics[width=0.8\columnwidth]{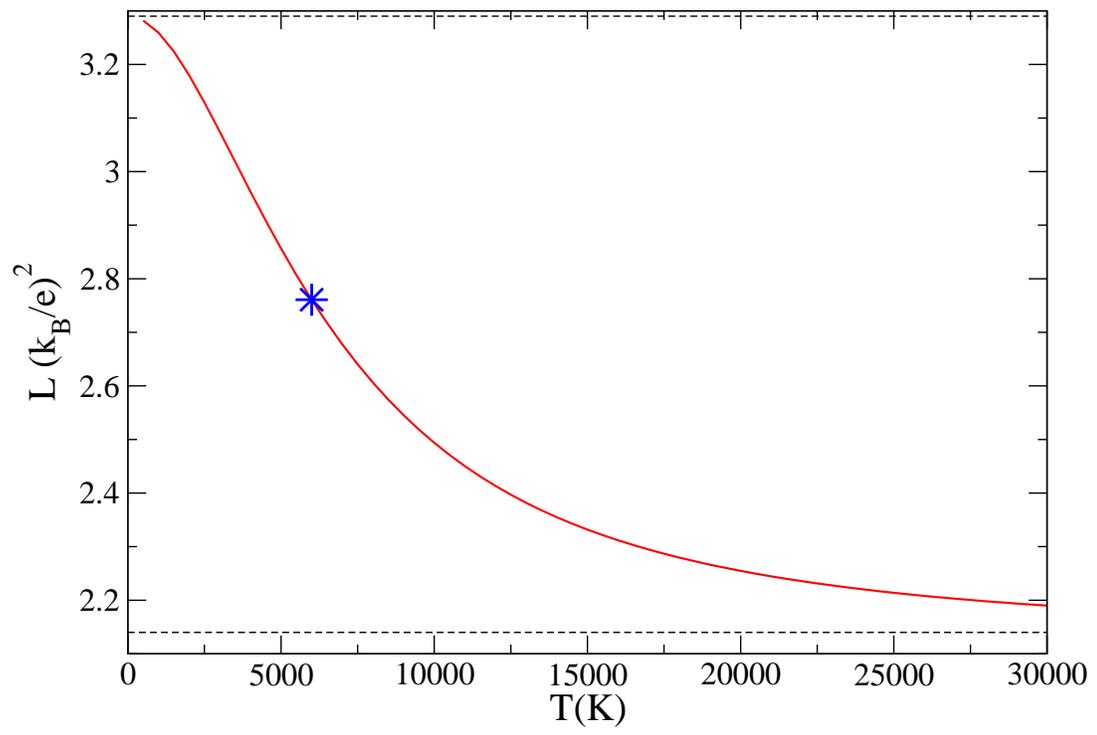}
 	\end{center}
 	\caption{\label{fig:L_vs_T} The Lorenz number vs. temperature. The star indicates the value of $L$ at the inner core temperature of 6000~K.
 	}
 \end{figure}

The Lorenz number depends on the magnitude of the electron-electron scattering compared to that of the other scattering processes. As an illustration of this, we consider elastic, temperature independent
scattering whose magnitude we set to the $\tau_{e-ph}$ at 6000~K and plot the
Lorentz number as a function of temperature in Fig.\ref{fig:L_vs_T}. At high temperatures the
electron-electron scattering that in a Fermi liquid increases quadratically with temperature dominates and the Lorentz number approaches the pure Fermi liquid result of 2.14~$(k_B/e)^2$. At low temperatures the electron-electron scattering is insignifcant and standard Lorenz number of $\pi^2/3 (k_B/e)^2$ is recovered
instead.

\end{document}